# Nonlinear shift along the sensorimotor-association-axis in brain responses to task performance

Yuqi Yuan[1], Bohan Zhang[2], Kyle Perkins[3], Xiaohui Yan[1], Fan Cao[1*]

[1] Department of Psychology, the University of Hong Kong
[2] School of Sciences, Chang'an University
[3] Florida International University

[*], correspondence author
Please address all correspondences to Dr. Fan Cao at fancao@hku.hk.

# Abstract

In the literature of cognitive neuroscience, researchers tend to assume a linear relationship between brain activation level and task performance; however, controversial findings have been reported in participants at different ages and different proficiency levels. Therefore, there may be a non-linear relationship between task performance and brain activation if a full range of task performance is considered. In the current study, using the Human Connectome Project (HCP) dataset we examined the relationship between brain activation and working memory performance in two conditions (i.e. faces and places). We found a gradual change from a U-shaped relationship to an inverted U-shaped relationship along the sensorimotor-association (S-A) axis in the face condition. In other words, in low-order sensorimotor areas, it is U-shaped and in the high-order prefrontal and association areas, it is inverted U-shaped, which suggests different properties in the encoding/representation region and in the cognitive calculation regions. However, in the place condition, such a shift is missing, presumably because most of the regions that are sensitive to task performance in the place condition are in the lower end of the S-A axis. Taken together, our study revealed a novel difference of functional property in response to task performance in the sensorimotor areas versus the association areas.

# 1 Introduction

Working memory (WM) is a cognitive system responsible for temporary maintenance and manipulation of information necessary for advanced cognitive tasks such as learning, reasoning and comprehension (Baddeley, 1992; D'Esposito and Postle, 2015). WM capacity increases with learning and practice (Asp et al., 2021; Brady et al., 2016; Jackson and Raymond, 2008; Xie and Zhang, 2017). With the increase of WM capacity over training, some studies found increased brain activities in the lateral prefrontal cortex (Olesen et al., 2004; Klingberg, 2010), while other studies found decreased activities (Buschkuehl et al.,2012, 2014). The inconsistency may be due to the different task difficulty and different range of WM capacity. Furthermore, traditional WM regions such as the dorsolateral prefrontal cortex (DLPFC) have been documented to often display an inverted U-shaped activation pattern with increasing WM load, reflecting a non-linear change of brain activation with WM demands (Callicott et al., 2003; Manoach, 2003; Van Snellenberg et al., 2015).

WM involves not only the DLPFC for cognitive control but also brain regions that directly represent the information that needs to be maintained in the WM task (Ranganath and D'Esposito, 2005; Rottschy et al., 2012; Carpenter et al., 2000; Nee et al., 2013; Sreenivasan et al., 2014). The cognitive control regions and the information representation regions may show different response patterns with increasing WM capacity. The lower-order information representation regions may show an increase with increasing WM (Booth et al., 2003) due to greater demands on information represented in that region, while the higher-order associative cortices that attune executive functions, on the other hand, may exhibit a non-linear change due to increasing effort followed by greater automatic processing (Persson et al., 2007; Parasuraman, 2000; Marois and Ivanoff, 2005; Vohs et al., 2018). However, no studies have systematically examined how different brain regions respond to increasing WM capacity.

The sensorimotor-to-association (S-A) axis is proposed to be an organizational principle of the human brain which captures the continuous gradient and hierarchical progression of anatomical, functional, evolutionary, and developmental properties of the cortex from primary sensorimotor to trans-modal association areas (Sydnor et al., 2021; Sydnor et al., 2023). Regions positioned at the low end of the S-A axis—such as the primary visual, auditory, somatosensory, and motor cortices—are highly myelinated, primarily involved in unimodal sensory and motor functions, and early developing (Figure 5a). In contrast, regions at the high end of the axis, the trans-modal association cortices such as the prefrontal and posterior parietal cortex are less myelinated, more plastic, late developing, and involved in abstract, sophisticated, and integrative processes such as mentalizing and executive control. Therefore, there may be a gradual change along the S-A axis in the brain response pattern to increasing WM capacity.

A recent training study examined brain changes with WM training of complex fractal patterns (Miller et al. 2022) and the researchers found that the mid-ventral PFC is involved in representing the stimuli with a non-linear curve as the WM capacity increases, suggesting that the lateral prefrontal cortex (LPFC) is not only involved in executive control but also in information representation itself. However, this study only examined the LPFC without concerning other brain regions, and only one type of stimuli was used in this study. Different types of stimuli may evoke different patterns in the brain because distinct neural resources are needed to represent and maintain these stimuli.

In the current study, we used the Human Connectome Project (HCP) dataset to investigate the relationship between brain activation and task performance on a 2-back WM task across two visual stimulus types: faces and places/scenes. A 2-back cognitive test involves a person identifying if the current stimulus matches one presented two trials ago. Specifically, we expected gradual change of the relationship with increasing WM capacity along the sensorimotor-association axis (S-A). We also expected selective responsiveness to faces and places in some brain regions. For example, faces require precise encoding of configurative features in the fusiform face area (FFA) (Kanwisher et al., 1997), while places involve the integration of spatial layouts and contextual information (Epstein, 2008) in the parahippocampal place area (PPA) (Epstein and Kanwisher, 1998). On the other hand, both faces and places may engage the LPFC for information maintenance (Ranganath et al. 2004). Taken together, our first aim is to examine whether the relationship between brain activation and WM capacity varies with the rank of S-A axis. The second aim is to examine whether the relationship depends on stimulus type.

## 2 Materials and Methods

### 2.1 Participants

The HCP included 1072 participants with complete fMRI and behavioral data in the WM task including the face and the place condition. From this original sample, we first removed 27 participants due to fMRI artifacts or low data quality. Then, if a participant's activation/deactivation was 4 standard deviations away from the group mean in more than 10% of the cortical parcellations, the participant was excluded (Face: $n = 11$, Place: $n = 10$). We also excluded participants whose behavioral performance on the task was 3 standard deviations away from the group mean (Face: $n = 10$, Place: $n = 14$). The final sample in each stimulus type is Face: $n = 1024$, age ranged from 22 to 37 years ($M = 28.80$, $SD = 3.71$), including 464 males (45.3%) and 560 females (54.7%); and Place: $n = 1021$, age ranged from 22 to 37 years ($M = 28.76$, $SD = 3.68$), including 470 males (46.0%) and 551 females (54.0%).

## 2.2 Task Paradigm

There were two scans for the WM task with two runs in each scan. In each run, there were eight task blocks and four fixation blocks. In four of the task blocks, it was a two-back working memory task, which required the participant to press a "Yes" button with an index finger when the current stimulus matched the one presented two trials earlier, and a "No" button when it did not match with a middle finger. The other four task blocks had a 0-back working memory task. Each block contained one of the four types of stimuli: faces, places, tools, and body parts. We only used data from the 2-back WM task to avoid a ceiling effect in the task performance. We also excluded data from the tools and body parts to simplify the results. There were 10 trials of 2.5s each, totaling 25s in each task block. Therefore, each stimulus type (i.e. faces and places) had 40 trials in total.

A 2.5s cue at the beginning of each block indicated the task type (2-back or 0-back). During each trial, the stimulus was presented for 2s, followed by a 500ms inter-trial interval (ITI). Each block contained approximately 2 yes-trials, 2–3 no-trials, where the stimulus was a repeat but not in the 2-back position, designed to test for false positives, and 5–6 no-trials which were new or not matching any recent stimulus, serving to maintain the task's demand on working memory.

**Image Acquisition**

All images were acquired using a customized Siemens 3T "Connectome Skyra" scanner with a standard 32-channel head coil. Whole-brain task fMRI data were acquired using a multi-band echo-planar imaging (EPI) sequence with the following parameters: repetition time (TR) = 720 ms, echo time (TE) = 33.1 ms, flip angle = 52°, bandwidth (BW) = 2290 Hz/Px, in-plane field of view (FOV) = 208 × 180 mm, 72 slices, and 2.0 mm isotropic voxel resolution. A multi-band acceleration factor of 8 was applied to enhance acquisition efficiency.

Spin echo phase-reversed images were acquired during the fMRI sessions, which were used to align T2-weighted (T2w) and fMRI images to the T1-weighted (T1w) structural images for each participant. Structural images were acquired with the following parameters: TR = 2400 ms, TE = 2.14ms, flip angle = 8°, BW = 210 Hz/Px, FOV = 224 × 224 mm, 256 slices and 0.7 mm isotropic voxel resolution for T1w images; TR = 3200 ms, TE = 565ms, variable flip angle, FOV = 224 × 224 mm, 256 slices and 0.7 mm isotropic voxel resolution for T2w images.

**2.3 fMRI Data Analysis**

For the analysis of the task fMRI data, we utilized the preprocessed and individual analyzed data provided by the HCP dataset. For our analysis, we used the contrast of 2-back of the face and place condition minus fixation baseline from the general linear model (GLM) to examine how brain activation changes with WM task performance.

We performed parcel-wise analyses using the HCP multi-modal parcellation 1.0 atlas (Glasser et al., 2016), including 180 distinct cortical parcels per hemisphere, and 360 in total. Specifically, we used the HCP-provided Workbench commands to parcellate the cortical surface and calculate the average parameter estimate for each stimulus type in the 2-back WM task in each parcel. This process generated a map of average parameter estimates for 360 parcellations per participant. These estimates, representing the brain activation level at each parcel, were subsequently used as input for the generalized additive model (GAMs) to fit a response curve with task performance in each parcel.

**2.5 Task performance**

Task performance was measured with Balanced Integration Score (BIS), which was defined as standardized accuracy minus standardized response time (RT) (Liesefeld et al., 2015; Liesefeld and Janczyk, 2019). We used z-score as the standardized score for accuracy and RT. The z-score was calculated by subtracting the group mean from the participant's value which was then divided by the standard deviation.

BIS can effectively mitigate the effect of speed-accuracy trade-off (SAT) by integrating RT and accuracy with equal weight, which has been applied in the domain of psychology and neuroscience (Liesefeld and Janczyk, 2019). Using a drift-diffusion model (Ratcliff, 1978; Ratcliff et al., 2016), Liesefeld and Janczyk (2019) showed that BIS is more robust to SAT-induced distortions than other measures in a simulation study, as it accurately reflects the underlying drift rate while filtering out spurious effects. We also conducted a simulation and the results confirmed the robustness of BIS against SAT effects and its capability to reflect task performance (See supplementary material for the simulation results and justifications).

**Examination of the relationship between brain and performance**

Generalized Additive Models (GAMs) were calculated to characterize the non-linear relationship between brain activation and performance (measured by BIS). GAMs extend linear regression by modeling the response of a predictor (e.g., BIS) with a combination of smooth functions (e.g. splines), avoiding restrictive assumptions of linearity. This flexibility is crucial in examining brain–behavior associations which often exhibit complex, non-linear effects (Feldman, 2025, Sydnor et al., 2023). In order to examine how brain activation changes with WM task performance, for each parcel, we fit a GAM with BIS as the smooth term while controlling for gender and age:

$$\text{Cortical Parcel Activation} \sim s(BIS) + Gender + Age$$

Where $s(\cdot)$ indicates the smooth term in GAM.

To filter out those cortical parcels that significantly respond to task performance, we conducted analysis of variance (ANOVA) to compare the variance explained by the full model with that by a nested, reduced model which does not contain the BIS smooth term, using the default hypothesis test in the mgcv package (Wood, 2000; Wood, 2017 (pp. 304). The statistical significance was assessed by a chi-square test, and a significant result indicates that the deviance of residuals is significantly smaller when the smooth term of BIS is included in the model. We corrected for multiple comparisons across the 360 GAMs using the false discovery rate (FDR) correction, setting the significance level at $P_{FDR} < 0.05$.

**Definition of linear and non-linear response patterns**

We calculated the first-order derivative and second-order derivative for each curve and identified 6 types of relationships between brain activation and BIS: linear increase, linear decrease, concave increase, concave decrease, convex increase, and convex decrease (Figure 1). The mean first-order derivative indicates whether the brain activation increases or decreases with BIS, with a positive mean first derivative corresponding to increase, and a negative mean first derivative corresponding to decrease. The mean second derivative indicates whether the curve is linear or non-linear. If the curve's mean second derivative is within the range of -0.05 and 0.05, then the curve was defined as linear in the current study. If the mean second derivative is greater than 0.05, the non-linearity was defined as convex; if the mean second derivative is lower than -0.05, it was defined as concave.

**Correlation with the S-A axis**

Next, we calculated the correlation between the non-linearity of the curve and the S-A ranking of that parcel (Figure 4a) across all 360 parcels separately for faces and places. The non-linearity was indexed by the mean second derivative of each parcel's response curve. We applied the Gaussian Mixed Model (GMM) for automatic identification and removal of those near-zero mean second derivative values.

## 3. Results

### 3.1 Behavioral results

The mean accuracy is 89.6% (SD = 10.1) for the WM face task and 90.1% for the WM place task (SD = 9.5). The mean response time is 926 ms (SD = 172) for the WM face task and 924 ms (SD = 164) for the WM place task. We then calculated the BIS for each task condition and removed outliers that are 3 standard deviations away from each task condition's respective mean value (Face: $n$ = 10, Place: $n$ = 14). The resulting mean BIS value is 0.054 (SD = 1.452) for the face condition and 0.077 (SD = 1.384) for the place condition. Behavioral results are summarized in Figure 2. There is no significant difference between the two conditions in accuracy, RT or BIS in paired t-test (n = 1006; for accuracy: t = -1.501, p = 0.134; for RT: t = 0.077, p = 0.938; for BIS: t = -0.306, p = 0.760).

### 3.2 Relationship between brain activation and BIS

After the FDR correction, there are 101 parcels that showed a significant relationship with BIS in the place condition, and 91 parcels in the face condition (Figure S2, Table S3). We counted the number of regions showing each of the 6 patterns for the face and place condition (Table 1). Most regions showed linear increase; however, in the face condition, 28 regions showed concave increase and in the place condition, 27 regions showed convex increase. Figure 3a shows the 6 types of responses in the brain in each condition. Figure 3b shows regions with a non-linear relationship between brain activation and BIS after GMM selection. Furthermore, 19 regions showed the same type of relationship in the two conditions (Figure 4), and 23 regions showed different types of relationships (Figure 4). There are also 49 regions that only responded to BIS in the face condition and 59 regions that only responded to BIS in the place condition (Figure 4). (Table S3)

### 3.3 Correlation between the non-linearity and SA

We found a significant negative correlation between the non-linearity and S-A ranking for faces (160 regions with non-linearity after GMM selection, Spearman's rho: -0.475, $p < 0.001$; R-squared: 0.181), but not for places (138 regions with non-linearity after GMM selection, Spearman's rho: 0.122, $p = 0.153$; R-squared = 0.000356) (Figure 5b). The same pattern remains when all 360 regions were included without GMM selection (Figure S4a). We also found a negative correlation for the face condition (Spearman's rho: -0.484, $p < 0.001$; R-squared: 0.267) but not the place condition (Spearman's rho: -0.218, $p = 0.356$; R-squared: 0.111) when we only included regions with significant GAM fit with linear regions filtered out by GMM (Figure 5b). The same pattern remains when all significant regions were included without GMM selection (Figure S4e).

Then we calculated the correlation between the non-linearity and S-A ranking separately for increasing and decreasing regions. If the mean first derivative is positive, then it is increasing; if the first derivative is negative, then it is decreasing. We found for both increasing and decreasing regions, there is a negative correlation with S-A ranking in the face condition (Increasing regions: 112 regions after GMM selection, Spearman's rho: -0.537, $p < 0.001$, R-squared = 0.254; Decreasing regions: 48 regions after GMM selection, Spearman's rho: -0.312, p = 0.031; R-squared = 0.072) (Figure 5c). There was no correlation for the place condition no matter whether it is increasing or decreasing regions (Increasing regions: 138 regions after GMM selection, Spearman's rho: 0.130, p = 0.164, R-squared < 0.001; Decreasing regions: 22 regions after GMM selection, Spearman's rho: 0.133, p = 0.556; R-squared = 0.025) (Figure 5c). The same pattern remains when linear regions were not filtered out by GMM. (Figure S4b)

Last, we verified our findings by randomly dividing the sample into two groups and replicated the analysis. We found a negative correlation in the face condition in both sub-samples (Sample 1: 133 regions after GMM selection, Spearman's rho: -0.507, $p < 0.001$; R-squared = 0.258; Sample 2: 150 regions after GMM selection, Spearman's rho: -0.322, $p < 0.001$; R-squared = 0.090), and no correlation in the place condition in both samples (Sample 1: 82 regions after GMM selection, Spearman's rho: -0.072, p = 0.519; R-squared = 0.0018; Sample 2: 157 regions after GMM selection Spearman's rho: 0.075, p = 0.353; R-squared = 0.00058) (Figure 5d). When all regions were included without GMM selection, the same pattern remains (Figure S4c).

### 3.3 Comparison between the face and place condition

To examine why there is a negative correlation with S-A ranking only in the face condition but not the place condition, we compared the S-A ranking of parcels that showed a significant relationship with BIS in the GAM in the two conditions with a label-shuffling permutation procedure. We first pooled all performance-responsive parcels from both conditions into a single set, and then repeatedly (10,000 times) drew two random samples from this pool, with the sample size matched with the number of significant parcels in the two conditions. For each permutation, we computed the mean difference in S-A ranking in the two conditions, thus generating a null distribution of S-A ranking differences of regions between the two conditions. We found that the face-responsive regions had a higher S-A rank than the place-responsive regions ($p < 0.001$, Figure S3a).

The permutation test also showed that the face condition had significantly more responsive regions with a concave pattern (i.e. mean second derivative smaller than -0.05) (n = 37) than the place condition (n = 4), permutation $p < 0.001$ (Figure S3b). In contrast, more parcels showed a convex increase in the place condition (n = 31) than in the face condition (n = 7), permutation $p < 0.001$ (Figure S3b).

## 4. Discussion

### 4.1 Non-linear responses with BIS

We found significant non-linear responses with BIS in both the face and the place condition, even though most regions showed a linear response with BIS. Moreover, the non-linear response tends to be concave in the face condition and convex in the place condition. Last, in the face condition, we found a negative correlation between the non-linearity and S-A ranking with higher-ranking parcels showing greater concave responses than lower-ranking parcels, whereas there is not such a shift along the S-A axis in the place condition. To our knowledge, this is the first study showing a gradual change in the non-linear relationship between brain activation and task performance along the S-A axis.

Our results indicate that the concave response pattern may be a functional hallmark of the high-order association cortices, which is characterized by increasing activation at the moderate levels of WM capacity followed by decreasing activation at the high levels, reflecting the dynamic allocation of cognitive control resources. This is consistent with the classic capacity theory of the prefrontal cortex (Braver et al., 1997), which posits that as individuals become more proficient, activation in cognitive control regions reduces, because the neural efficiency increases (Haier et al., 2009; Neubauer & Fink, 2009). This theory explains why association regions such as the dorsomedial prefrontal cortex (DMPFC) and the dorsolateral prefrontal cortex (DLPFC) show peak activity at the intermediate performance. From low performance to intermediate performance, one needs to make more effort in order to increase the performance. However, for the high performers, brain activation decreases due to great automaticity, leading to the inverted-U response profile as observed in our results. We also found that the turning point is different in different regions (Figure 5a). Some regions are around 0, but there are also regions with a turning point above 0, suggesting that these regions reach saturation at a different performance level. However, the concave pattern in the high-order regions only exist in the face condition.

In contrast, the place condition is associated with a convex pattern in the lower-order regions. These regions are often responsible for encoding and maintaining perceptual features (e.g., spatial layouts or facial components) (Riesenhuber & Poggio, 1999). Convex responses suggest that the average performance is associated with the lowest activation. For low performers, these regions need to work harder to encode the visual perceptual information, and as performance increases, the brain activation decreases in these perceptual regions. However, when the performance surpasses the average and becomes even higher, more fine-tuned representation and sophisticated perceptual strategies may be involved, which leads to greater activation of these regions. The difference between the face and the place condition may be because more complex calculation is needed to discriminate faces than places, since faces share a greater visual similarity than places.

Furthermore, there is a gradual change to greater concave along the S-A axis in the face condition but not the place condition, underscoring the stimulus-specific nature of the effect. Our analysis suggests that this is because the place-related performance effects are largely confined to the unimodal and early multimodal areas (e.g., parietal and occipital cortex). Therefore, there is a lack of gradual change with S-A axis. Future research needs to investigate whether the non-linear shift pattern found in the face condition can be replicated using other stimuli for which the higher-rank regions are responsive to task performance.

We also examined the inflection points (the BIS value corresponding to the sign's change of first order derivative, e.g., the point where a concave curve ends increasing and starts decreasing along the performance-axis) of concave regions and convex regions from both datasets (Table 2). There are 28 concave regions and 2 convex regions from Face condition and 1 concave region and 15 convex regions from Place condition that have inflection points. Result shows that for all 29 concave regions versus all 17 convex regions, there is a significant difference between mean inflection point: concave regions' mean inflection point: BIS = 0.259, SD = 0.716; convex regions' mean inflection point: BIS = -1.17, SD = 1.07. Also, for the 28 concave face regions and the 15 convex place regions with inflection points, there exists a significant mean difference: concave face regions' mean inflection point: BIS = 0.238, SD = 0.720; convex place regions' mean inflection point: BIS = -0.951, SD = 0.811; permutation $p < 0.001$. Such significant result also holds true for the 26 concave increasing face regions and the 10 convex increasing regions: concave increasing face regions' mean inflection point: BIS = 0.379, SD = 0.517; convex increasing place regions' mean inflection point: BIS = -1.399, SD = 0.557; permutation $p < 0.001$. The permutation test between concave regions (n = 28) and convex regions (n = 2) within Face condition is significant: concave regions' mean inflection point: BIS = 0.238, SD = 0.720; convex regions' mean inflection point: BIS = -2.821, SD = 1.738; permutation $p$ = 0.0053. On the other hand, permutation test between convex regions (n = 15) and concave regions (n = 1) within Place condition show marginal significance: concave region inflection point: BIS = 0.840; convex regions' mean inflection point: BIS = -0.951, SD = 0.811; permutation $p$ = 0.062.

The inflection point, defined as the BIS score where the first-order derivative of the brain activation curve changes sign (e.g., from increasing to decreasing in concave curves; see Figure 6), indicates a pivotal threshold in the activation-performance relationship. Considering the shape of the curves specifically, the face's concave pattern (inverted U-shaped) had an inflection point (BIS > 0, above-average performance) that resulted in a higher threshold, indicating that activation only declined after performance increased substantially enough to warrant a relatively high level of performance. This finding indicates that automation and reduction of neural effort occurred later in the proficiency continuum: at lower-to-moderate BIS scores, activation was positively associated with performance because of increased cognitive effort, but once performance rode the average and into higher levels of performance, neural efficiency began to take hold, which resulted in decreased activation as processes quickly became more automatic (Braver et al., 1997; Neubauer & Fink, 2009).

In contrast, the place's convex pattern (U-shaped) has a lower inflection point (BIS < 0, below average performance), meaning activation occurs earlier even before hitting average level learning effort. This indicates that extra neural demands are incurred at an early stage to counter initial challenges to reach moderate performance; activation appears to level out to a slight degree before the inflection point and then next increases noticeably towards good performance above average. Such a pattern might suggest that task demands are met more easily at lower levels, as they do not require an increase in activation at the onset but do require a greater escalation for better performance—perhaps to also involve less or more finely tuned strategies or deal with greater complications and not to make the task too cognitively easy (Parasuraman, 2000; Marois & Ivanoff, 2005).. If the increase occurred only above average level, it would imply that the task was too easy such that high levels of skill did not require additional additional effort.

By highlighting the intrinsic characteristics of curve shapes, our results showcase how brain activation shifts in a nonlinear manner to performance at different levels along the S-A axis. The late decline in concave curves indicates a sustained effort in a higher-order regions until later levels of fluency, while the beginning increase in convex curves indicates a pre-emptive recruitment of lower-order areas to achieve baseline level fluency. These findings are useful for understanding neural efficiency during a WM task. For example, training programs could target explicit performance levels where activation transitions occur to accomplish efficient cognitive resource use. Future research can examine these curve patterns in informational constructs so that curve representations could be generalized across tasks and potentially examined computationally in models that reflect an underlying theory of transitions from effort to automaticity.

## 4.2 Shared regions in the face and place conditions
### *Shared regions and same patterns*

There were 19 regions that showed the same response pattern in the face and place condition (Figure 4). Most of these regions showed a linear increase pattern, including regions in the sensorimotor cortex, visual stimuli representation pathway, and frontal cortex. These regions with linear increase in both conditions may be involved in low level visual perception and sensorimotor processing of the task.

*Shared regions but different patterns in the face and place condition*

Several regions in the premotor cortex, the intraparietal lobe and the posterior cingulate cortex showed linear increase in faces and convex increase in places. Convex increase in places means that the brain activation does not change much with performance increase at the low end, but it changes rapidly at the high end. This difference between faces and places may be because faces have greater similarity than places and it needs more neural calculation to discriminate faces than places. Therefore, for faces, the increase of performance needs proportional increase of brain activation even for low performers. For places, however, at the low end of performance, performance increase does not require much change in brain involvement, because it is a relatively easy task.

In contrast, the medial temporal cortex including the right hippocampus showed a convex increase in the face condition but a linear increase in the place condition. This pattern may point to different mnemonic strategies for faces and places because the medial temporal cortex in general is more responsive in the place condition than the face condition, and a broader medial temporal cortex including the bilateral entorhinal cortex showed responses to BIS in places but not faces. Therefore, this part of the brain may be especially important for places.

Lastly, the left medial prefrontal showed linear increase in places but concave increase in faces. The medial prefrontal areas are involved in personal experience and memory (Eustron et al., 2012; Wagner et al., 2012), which may be involved differently in faces and places.

**4.3 Limitations**

This work has several limitations. First, the neurophysiological explanation of non-linear relationship with task proficiency is still not available. Why do the high rank regions show a concave pattern while the lower rank regions show a convex or linear pattern? Second, we only observed the systematic change with S-A rank in faces but not places. What are the key variables of tasks that determine the pattern? Future research is needed to examine other tasks and explain the generality.

**5. Conclusions**

The S–A gradient perspective (sensory–association axis) provides a useful framework: lower rank regions tend to show a convex curve while higher rank regions tend to show a concave curve. Our study advances brain–behavior mapping by examining how the relationship systematically changes along the S-A axis. In broader terms, these results support a view of the cortex as a multi-level processor. At the lower and intermediate level, sensory cortex (occipital, parietal) extracts visual features of faces or places and category-specific association areas (FFA, fusiform face area, PPA, parahippocampal place area) represent complex stimulus templates. At higher levels, frontoparietal regions play roles in the allocation of cognitive resources like attention and manipulate memory content. Our data show that face and place WM engage these levels differently: WM face leans on face-selective FFA, OFA (occipital face area) and some temporal regions for stimulus representation, frontal regions with distinctive concave patterns for task engagement and putative resource allocation, while WM place relies on ventromedial visual stream regions and

PPA for stimulus representation and involves distributed regions within premotor, intraparietal networks with distinctive convex patterns for putative attention maintenance and task outcome alert.

Overall, this study contributes to understanding the brain's functional organization. It suggests that the brain's macroscale gradients not only organize regions by functional specialization but also shape the response dynamics with task proficiency. By revealing the systematic mapping between gradient position (S–A rank) and response feature to behavioral proficiency, we demonstrate a novel principled mapping from brain topology to brain function. Such mapping is crucial for understanding how distributed brain systems give rise to working memory performance.

Table 1: Descriptive counts of six model curve response patterns observed in the cortical parcels across two visual stimuli (Place, Face)

| Response Pattern | Place | Face |
|---|---|---|
| Linear Increase | 64 | 46 |
| Linear Decrease | 2 | 1 |
| Concave Increase | 3 | 28 |
| Concave Decrease | 4 | 1 |
| Convex Increase | 27 | 6 |
| Convex Decrease | 1 | 9 |
| **Total** | **101** | **91** |

Table 2: Mean Inflection Point (BIS) of Significant GAM Regions' Curve, Separated by Region Type and Condition

|  | Face | Place |
|---|---|---|
| Concave | 0.238 (n = 28; SD: 0.720) | 0.834 (n = 1; SD: nan) |
| Convex | -2.821 (n = 2; SD: 1.738) | -0.951 (n = 15; SD: 0.811) |

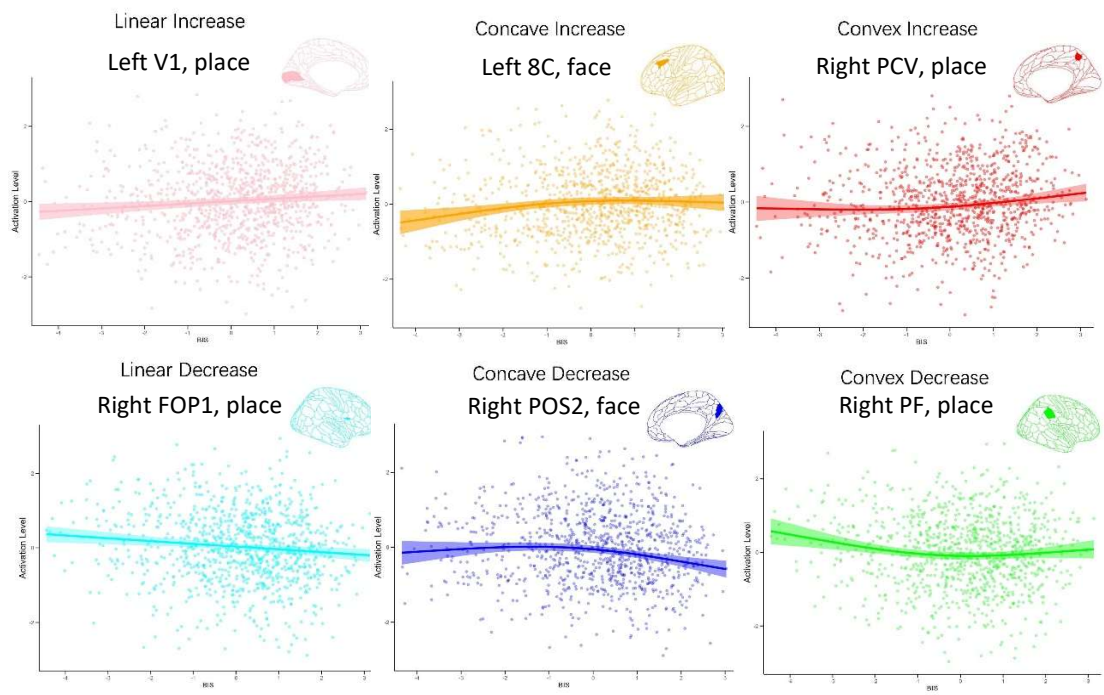

Figure 1: Region type sample plot visualization: normalized brain activation value response curve are overlaid with data from participants. Regional response curve GAM-predicted brain activation value at different performance (BIS) level, with a 95% credible interval band. Color of the plot and scatter is paired with the corresponding region type in Figure 3. The brain activation value is normalized for visualization purpose.

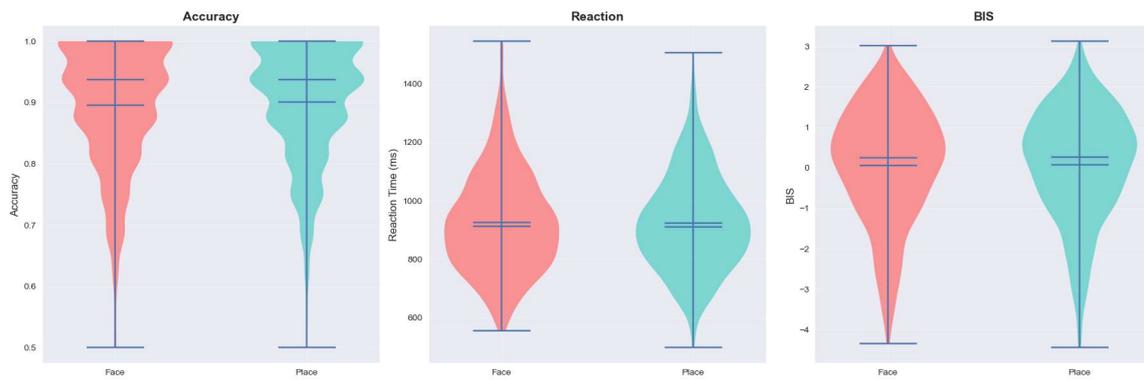

Figure 2: Violin graphs for accuracy, reaction time (RT) and balanced integration score (BIS) in the WM 2-back Face and WM 2-back Place tasks.

A. GAM significant regions

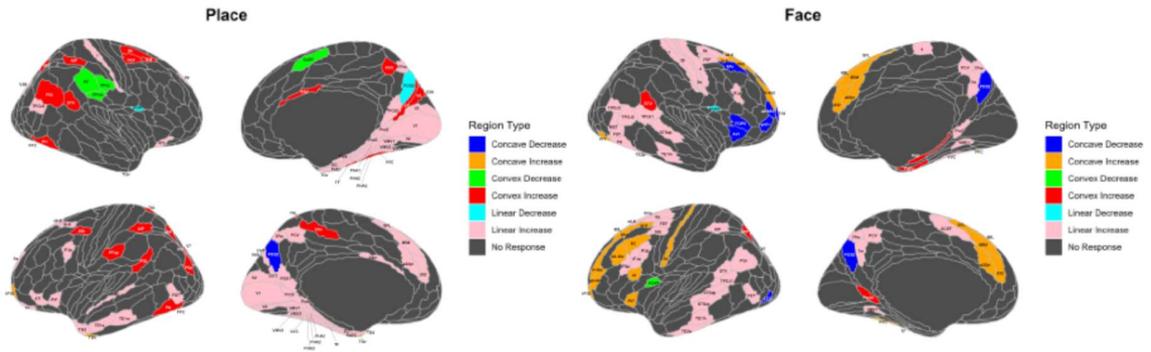

B. Regions with a non-linear relationship with BIS

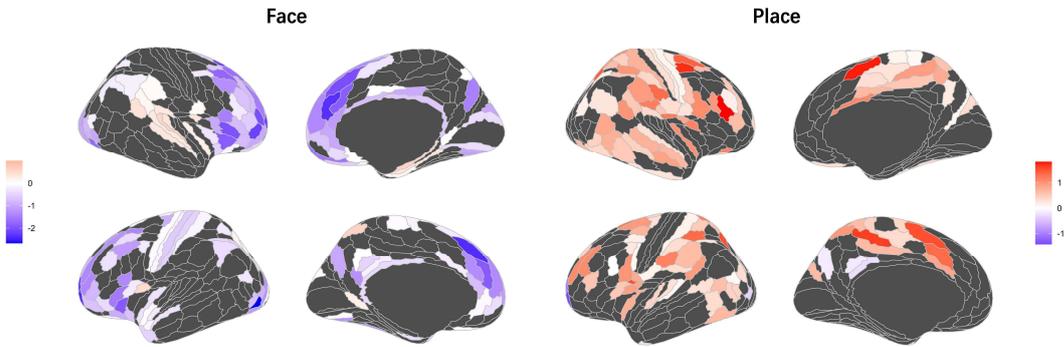

Figure 3: (**a**) Region type map of regions with significant GAM fit after FDR correction, separated by task stimuli: pink indicates linear increase pattern, cyan indicates linear decrease pattern; red indicate convex increase pattern, green indicate convex decrease pattern; orange indicate concave increase pattern, blue indicate concave decrease pattern. (b). Regions with a non-linear relationship with BIS. Color indicates the mean second derivative value of the region's GAM curve.

A. Same regions same patterns

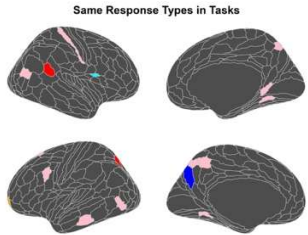

B. Same regions different patterns

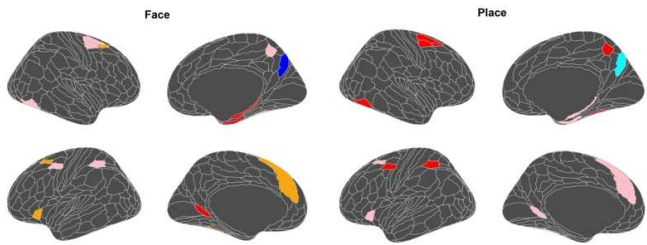

C. Different regions different patterns

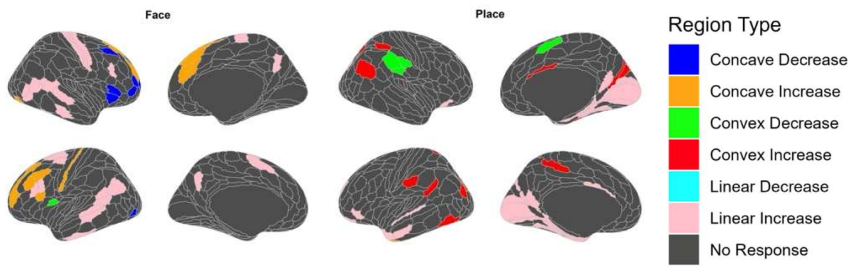

Figure 4. (a) same regions-same patterns: shared regions that show the same response pattern across two task conditions. (b) same regions-different patterns: shared regions that show different response patterns across the two task conditions. (c) Regions that respond to BIS only in the face or place condition.

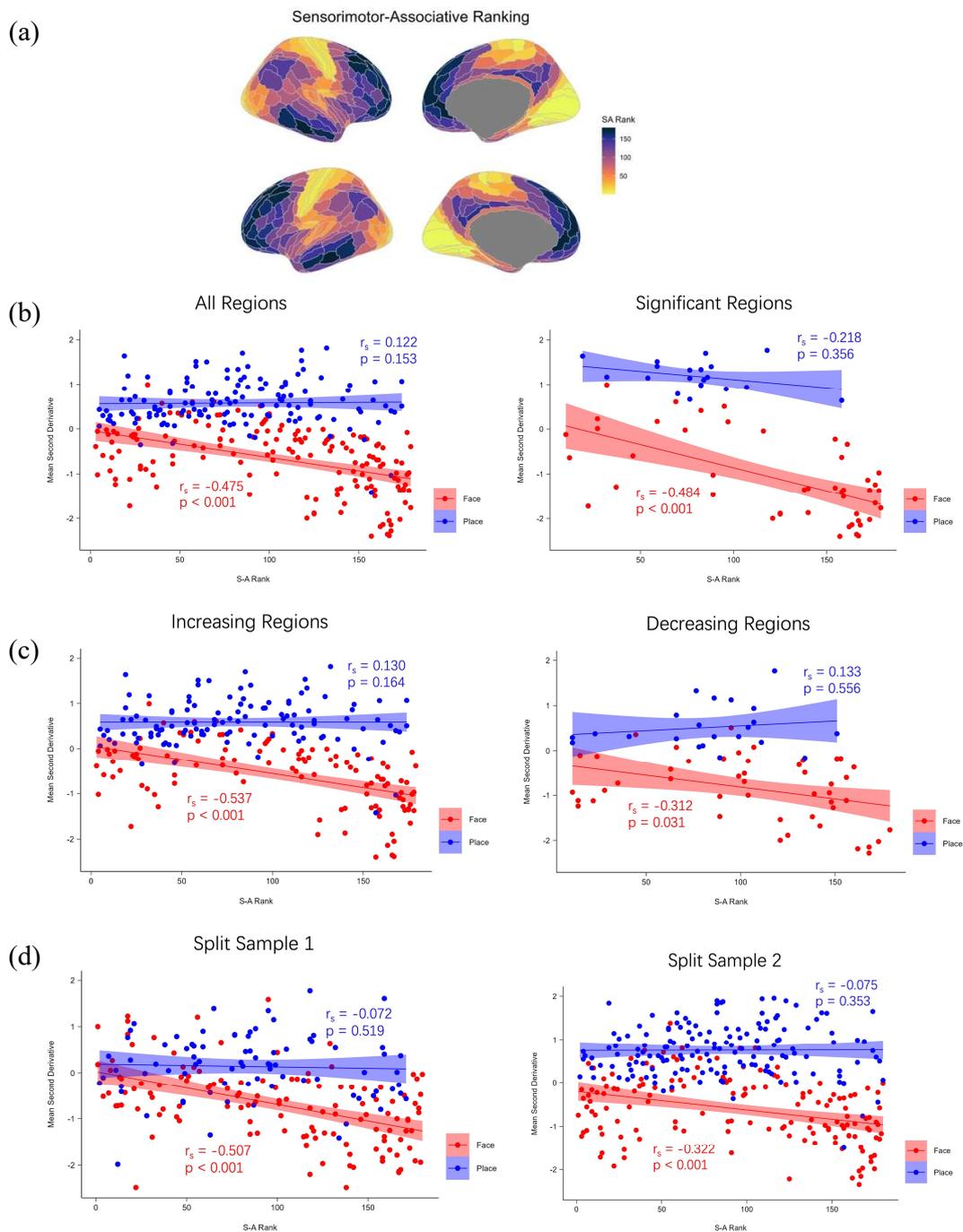

Figure 5. Relationship between cortical hierarchy (S-A ranking) and non-linear activation patterns in each stimulus type. Gaussian Mixed Model is used to filter out regions with trivial non-linearity. (a) Surface projection of S-A ranking across cortical parcels. Darker colors indicate higher association-level ranking. (b) Left: significant negative correlation between non-linearity and S-A ranking is observed in the face task, but not in the place task. Right: significant negative correlation between non-linearity and S-A ranking is observed in regions with significant GAM fit after false discovery rate (FDR) correction in the face task, but not in the place task. (c) Significant negative correlation between non-linearity and S-A ranking is observed in the face task in both regions with positive mean first derivative and negative mean first derivative, but not in the place task. (d) Two random split sample with the same sample sizes for WM face, both has significant relationship between SA rank and mean second derivative. Two random split sample with the same sample sizes for WM place, both fail to reach .001 level of significance under spearman regression coefficient.

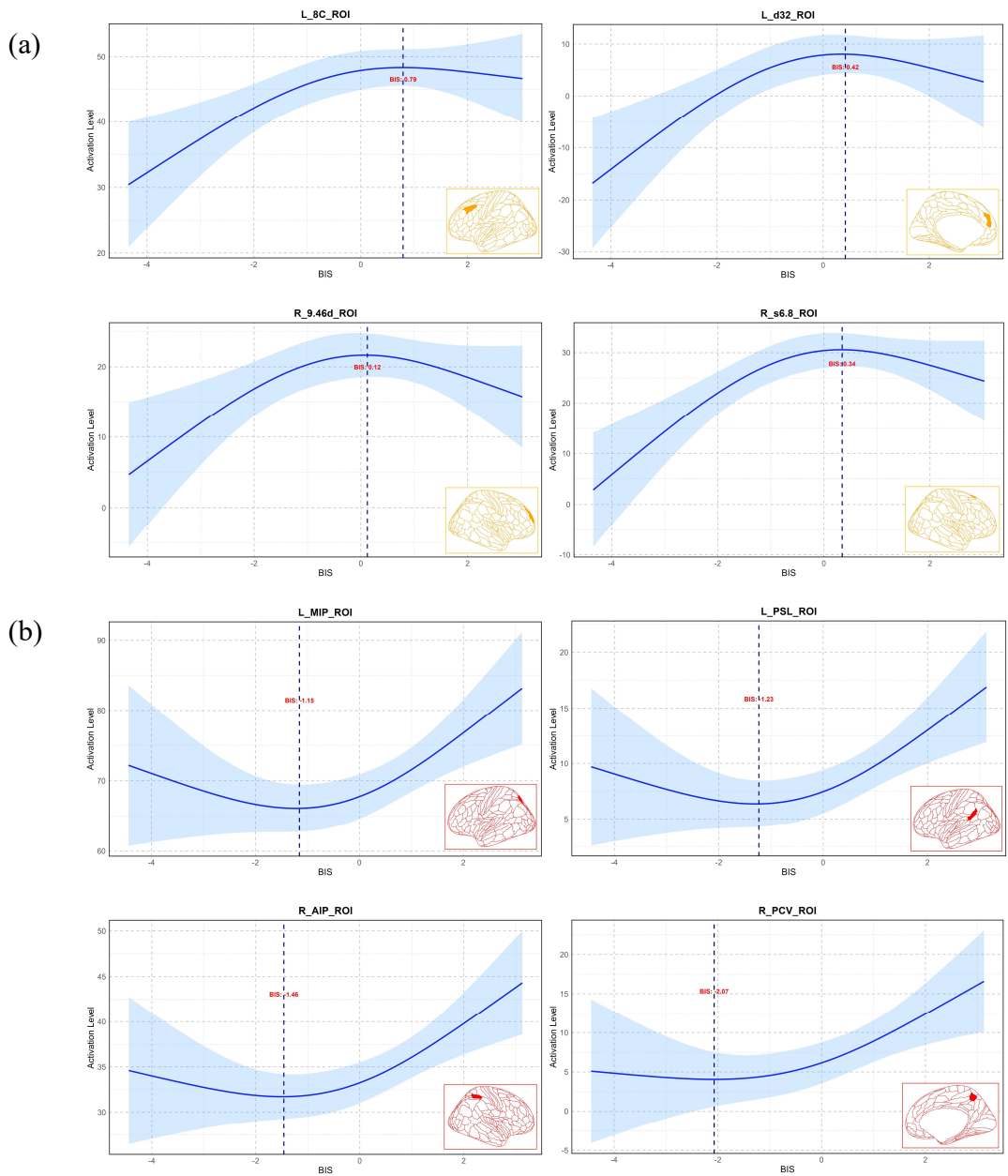

Figure 6: Concave and convex pattern sample plots, where the vertical dotted line indicates the turning point in each curve. (a)Concave patterns across the dorsal lateral prefrontal regions for face stimulus. (b) Convex patterns in distributed cortical regions for place stimulus

# Supplementary materials

Liesefeld & Janczyk (2019) defined BIS in terms of mean RT. Specifically, they defined BIS as

$$z_{PC} - z_{RT}$$

where $z_{PC}$ is the z-score of the percentage of correct responses, $z_{RT}$ is the z-score of mean RT conditional on correct responses. Note that $z_{RT}$ is standardized based on the standard deviation of mean RTs across participants and conditions, not across trials. However, we define our BIS in terms of median RT. We do this for two reasons: (a) data availability, since HCP only includes median RT, and (b) to account for skewed RT distributions and reduce the influence of outliers. We replicated Tables 1 and 2 from Liesefeld & Janczyk (2023) to show that the metric remains a robust, composite measure of speed and accuracy even defined in terms of median RT instead of mean RT. Our simulation method follows closely that of Liesefeld & Janczyk (2023): In Table 1, we simulated two sets of data, one with a variation in drift rate $v$ ("real" effect) and one with a variation in threshold separation $a$ (SAT effect). We model the decision process as an arithmetic Brownian motion, that is

$$dX_t = v\, dt + \sigma\, dW_t.$$

To normalize, we set $\sigma = 4$. A decision is made when $X_t$, starting at $X_0 = a/2$, exceeds either the upper threshold $a$ (correct response) or the lower threshold $0$ (incorrect response). Time spent on additional processes of encoding and responding is captured via an additional non-decision time parameter, $t^{ER}$, which is added to the decision time to yield the overall RT. To simulate natural variability across different individuals and trials, two random terms are added to each parameter:

$$\mu_{i,j} = \mu_j + \varepsilon_i^{between} + \varepsilon_{i,j}^{within}.$$

The error terms $\varepsilon_i^{between}$ and $\varepsilon_{i,j}^{within}$ were drawn from a set of random variables $E^{between} \sim N(0, \sigma_B^2)$ and $E_j^{within} \sim N(0, \sigma_W^2)$, respectively. For the drift rate simulation, we set $\sigma_B^2 = 0.01^2$ and $\sigma_W^2 = 0.005^2$; for the SAT simulation we set $\sigma_B^2 = 20^2$ and $\sigma_W^2 = 10^2$. The non-decision time $t_i^{ER} \sim N(300, 20)$ was drawn separately for each participant $i$, but was the same for across conditions.

In addition, we conducted simulations to evaluate the sensitivity and robustness of the BIS defined with median RT. These simulations systematically varied the drift rate parameter, which influences both reaction time and accuracy in cognitive tasks. Figure 1(a) illustrates the percentage of significant comparisons at p<0.05 between v0 and a variable drift rate, depicted as the horizontal axis on the graph. Figure 1(b) complements this by showing the corresponding effect sizes, providing insight into the magnitude of the differences captured by the BIS metric at different drift rates. In Table 2, we varied the threshold separation a while keeping v at different levels. Each simulation contains 1000 experiments with 20 participants each and 1000 trials per condition.

Table S1: Effects on different measures

| Measure | Mean 1 | Mean 2 | Effect size $d_z$ | % significant |
|---|---|---|---|---|
| **"Real" effect** ($v_1 = 0.246$ vs. $v_2 = 0.261$) | | | | |
| Mean RT | 455 | 451 | 0.52 | 56.0 |
| PC | 0.88 | 0.89 | −0.82 | 92.8 |
| BIS | −0.51 | 0.51 | −0.92 | 96.9 |
| **SAT effect** ($a_1 = 120$ vs. $a_2 = 130$) | | | | |
| Mean RT | 446 | 464 | −0.68 | 78.9 |
| PC | 0.87 | 0.89 | −0.57 | 66.0 |
| BIS | −0.009 | 0.009 | −0.04 | 4.0 |

Table S2: Additional simulations with SAT effects

| Measure | Mean 1 | Mean 2 | Effect size $d_z$ | % significant |
|---|---|---|---|---|
| **Case 1:** $v = 0.35$, $a_1 = 110$ vs. $a_2 = 130$ | | | | |
| Mean RT | 410.00 | 437.00 | −1.40 | 100.0 |
| PC | 0.92 | 0.94 | −1.08 | 99.5 |
| BIS | −0.028 | 0.028 | −0.11 | 6.6 |
| **Case 2:** $v = 0.35$, $a_1 = 115$ vs. $a_2 = 125$ | | | | |
| Mean RT | 417 | 430 | −0.68 | 79.4 |
| PC | 0.93 | 0.94 | −0.56 | 65.2 |
| BIS | −0.018 | 0.018 | −0.07 | 4.9 |
| **Case 3:** $v = 0.11$, $a_1 = 200$ vs. $a_2 = 220$ | | | | |
| Mean RT | 736.00 | 809.00 | −1.30 | 100 |
| PC | 0.80 | 0.82 | −0.92 | 95.6 |
| BIS | 0.03 | −0.03 | 0.08 | 5.1 |
| **Case 4:** $v = 0.11$, $a_1 = 205$ vs. $a_2 = 215$ | | | | |
| Mean RT | 754.00 | 790.00 | −0.65 | 74.8 |
| PC | 0.81 | 0.82 | −0.45 | 44.5 |
| BIS | 0.02 | −0.02 | 0.06 | 4.6 |
| **Case 5:** $v = 0.11$, $a_1 = 110$ vs. $a_2 = 130$ | | | | |
| Mean RT | 453.00 | 505.00 | −1.36 | 100 |
| PC | 0.69 | 0.72 | −1.02 | 98.7 |
| BIS | 0.013 | −0.013 | 0.04 | 5.0 |

Note that in Table S2, the percentage of significant simulations is around 5%, confirming that the BIS with median RT doesn't inflate empirical p-values, unlike the BIS with mean RT.

Table S3: Region's Response Pattern

| Label | Face Response Pattern | Place Response Pattern |
|---|---|---|
| **Shared-Same Regions** | | |
| Right 7Pm | linear increase | linear increase |
| Right TPOJ3 | linear increase | linear increase |
| Left TF | linear increase | linear increase |
| Left FST | linear increase | linear increase |
| Left PCV | linear increase | linear increase |
| Left 7Pm | linear increase | linear increase |
| Left s6-8 | linear increase | linear increase |
| Left TE1m | linear increase | linear increase |
| Right 1 | linear increase | linear increase |
| Right ProS | linear increase | linear increase |
| Right VMV1 | linear increase | linear increase |
| Left IFJp | linear increase | linear increase |
| Left VMV1 | linear increase | linear increase |
| Left a10p | concave increase | concave increase |
| Left 7PL | convex increase | convex increase |
| Left MIP | convex increase | convex increase |
| Right STV | convex increase | convex increase |
| Right FOP1 | linear decrease | linear decrease |
| Left POS2 | concave decrease | concave decrease |
| **Total: 19** | | |
| **Shared-Changed Regions** | | |
| Left 8BM | concave increase | linear increase |
| Left AVI | concave increase | linear increase |
| Left SFL | concave increase | linear increase |
| Left a32pr | concave increase | linear increase |
| Left d32 | concave increase | linear increase |
| Right FFC | concave increase | linear increase |
| Left 8BL | concave increase | linear increase |
| Left i6-8 | concave increase | linear increase |
| Left VVC | concave increase | linear increase |
| Right V8 | concave increase | linear increase |
| Left AIP | linear increase | convex increase |
| Right 6a | linear increase | convex increase |
| Right PH | linear increase | convex increase |
| Right FEF | linear increase | convex increase |
| Right VVC | linear increase | convex increase |
| Left 55b | linear increase | convex increase |
| Right PCV | linear increase | convex increase |
| Left V7 | linear increase | convex increase |
| Right POS2 | concave decrease | linear decrease |
| Right i6-8 | concave increase | convex increase |
| Left ProS | convex increase | linear increase |
| Right H | convex increase | linear increase |
| Right PeEc | convex increase | linear increase |
| **Total: 23** | | |
| **Face-Only Regions** | | |
| Right MST | linear increase | NA |
| Right 7m | linear increase | NA |
| Right IFJa | linear increase | NA |
| Right STSdp | linear increase | NA |
| Right TE2p | linear increase | NA |
| Right TPOJ1 | linear increase | NA |
| Right TPOJ2 | linear increase | NA |

| Region | | |
|---|---|---|
| Right FST | linear increase | NA |
| Right TE1m | linear increase | NA |
| Left STV | linear increase | NA |
| Left 7m | linear increase | NA |
| Left IFJa | linear increase | NA |
| Left IFSp | linear increase | NA |
| Left STSdp | linear increase | NA |
| Left STSvp | linear increase | NA |
| Left TPOJ1 | linear increase | NA |
| Left PGi | linear increase | NA |
| Left TE2a | linear increase | NA |
| Right 4 | linear increase | NA |
| Right 3b | linear increase | NA |
| Right 3a | linear increase | NA |
| Left FEF | linear increase | NA |
| Left SCEF | linear increase | NA |
| Left 6ma | linear increase | NA |
| Left 6a | linear increase | NA |
| Left LO2 | concave decrease | NA |
| Right PIT | concave increase | NA |
| Right SFL | concave increase | NA |
| Right d32 | concave increase | NA |
| Right 8BM | concave increase | NA |
| Right a32pr | concave increase | NA |
| Left 8C | concave increase | NA |
| Left 44 | concave increase | NA |
| Right 8Ad | concave increase | NA |
| Right 8BL | concave increase | NA |
| Right 9-46d | concave increase | NA |
| Right s6-8 | concave increase | NA |
| Left p9-46v | concave increase | NA |
| Left a9-46v | concave increase | NA |
| Left 9-46d | concave increase | NA |
| Left 3a | concave increase | NA |
| Left 8Ad | concave increase | NA |
| Right AVI | concave decrease | NA |
| Right FOP5 | concave decrease | NA |
| Right 8Av | concave decrease | NA |
| Right a9-46v | concave decrease | NA |
| Right p10p | concave decrease | NA |
| Right p47r | concave decrease | NA |
| Left FOP3 | convex decrease | NA |
| **Total: 49** | | |
| **Place-Only Regions** | | |
| Left 33pr | NA | linear increase |
| Left PHA2 | NA | linear increase |
| Left 47l | NA | linear increase |
| Left 9a | NA | linear increase |
| Left 11l | NA | linear increase |
| Left EC | NA | linear increase |
| Left A5 | NA | linear increase |
| Left PeEc | NA | linear increase |
| Left DVT | NA | linear increase |
| Left PHA1 | NA | linear increase |
| Left VMV3 | NA | linear increase |
| Left V6A | NA | linear increase |
| Left TGd | NA | linear increase |
| Left TE1a | NA | linear increase |
| Left PreS | NA | linear increase |
| Left PHA3 | NA | linear increase |
| Right V1 | NA | linear increase |
| Left V3A | NA | linear increase |
| Right V2 | NA | linear increase |
| Right V7 | NA | linear increase |
| Right V3B | NA | linear increase |
| Right POS1 | NA | linear increase |
| Right 7PL | NA | linear increase |
| Right 9p | NA | linear increase |
| Right 47s | NA | linear increase |
| Right EC | NA | linear increase |
| Left POS1 | NA | linear increase |
| Right PreS | NA | linear increase |
| Right PHA3 | NA | linear increase |
| Right VMV3 | NA | linear increase |
| Right PHA2 | NA | linear increase |
| Right VMV2 | NA | linear increase |
| Right TGv | NA | linear increase |
| Left V1 | NA | linear increase |
| Left VMV2 | NA | linear increase |
| Left V8 | NA | linear increase |
| Right PHA1 | NA | linear increase |
| Left V2 | NA | linear increase |
| Left TGv | NA | concave increase |
| Left FFC | NA | concave increase |
| Left PSL | NA | convex increase |
| Right PGi | NA | convex increase |
| Right DVT | NA | convex increase |
| Right TF | NA | convex increase |
| Left 5mv | NA | convex increase |
| Left PH | NA | convex increase |
| Right AIP | NA | convex increase |
| Left 7AL | NA | convex increase |
| Left PGp | NA | convex increase |
| Right 33pr | NA | convex increase |
| Right V6 | NA | convex increase |
| Right MIP | NA | convex increase |
| Left IP0 | NA | convex increase |
| Left PFop | NA | convex increase |
| Right V3A | NA | convex increase |
| Right PF | NA | convex decrease |
| Right PFop | NA | convex decrease |
| Right SCEF | NA | convex decrease |
| Right PFcm | NA | convex decrease |
| **Total: 59** | | |
| **Grand Total:** | **150** | |

Table S3: Region's response pattern table. Regions with significant GAM fit after FDR correction are classified into four classes based on their response pattern in face and place condition. Shared-same regions show significant GAM fit in both conditions, and their response patterns are the same across the two conditions. Shared-changed regions also show significant GAM fit in both conditions, but their response patterns differ between the two conditions.

Figure S1:

(a)

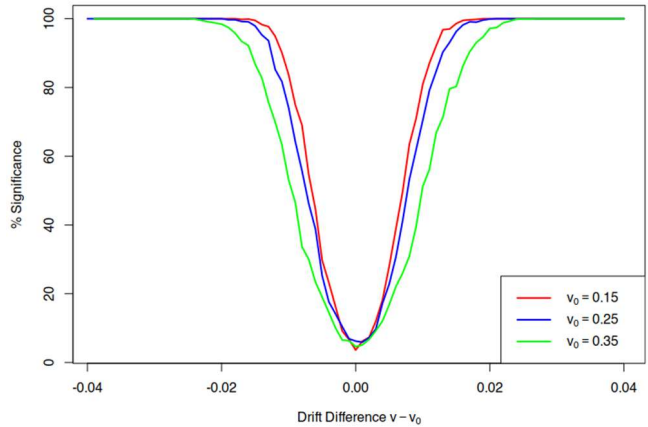

(b)

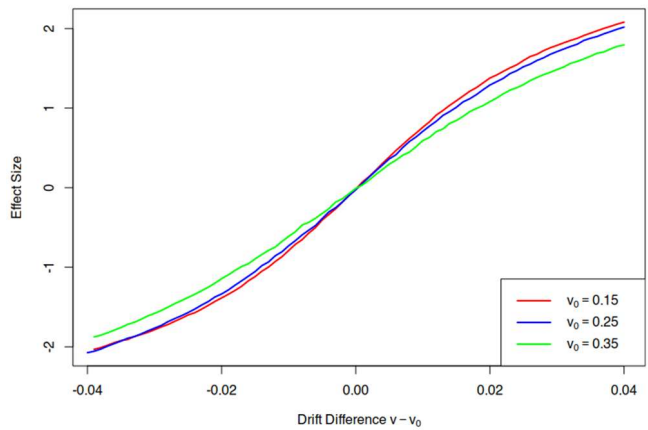

Simulation results for the sensitivity of the BIS defined with median RT. Panel (a) shows the percentage of significant comparisons at p<0.05 (statistical power), while Panel (b) presents the corresponding effect sizes (difference in BIS). The results show that BIS is sensitive to, and has an approximately linear relationship w.r.t. the underlying drift parameter.

Figure S2:

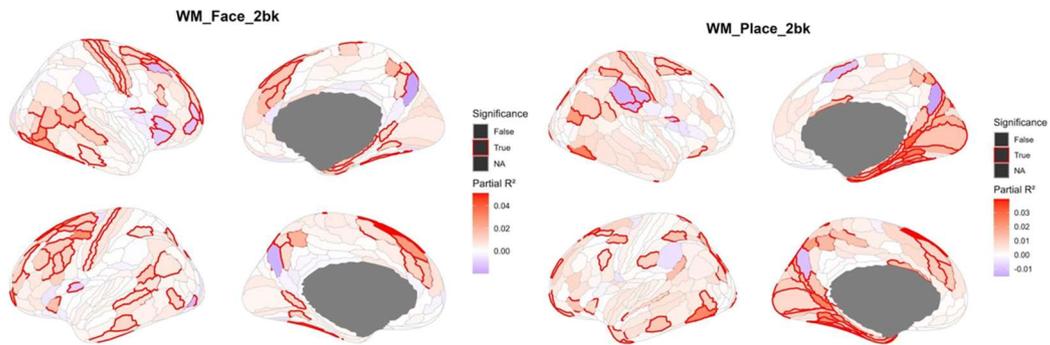

Effect size map. Regions with a red borderline showed a significant relationship with BIS after FDR correction. The color indicates the Partial R square of the full GAM model with BIS compared to a null model with only covariates. The sign of effect size is determined by the direction of response curve's mean first derivative for visualization.

Figure S3:

(a)

(b)

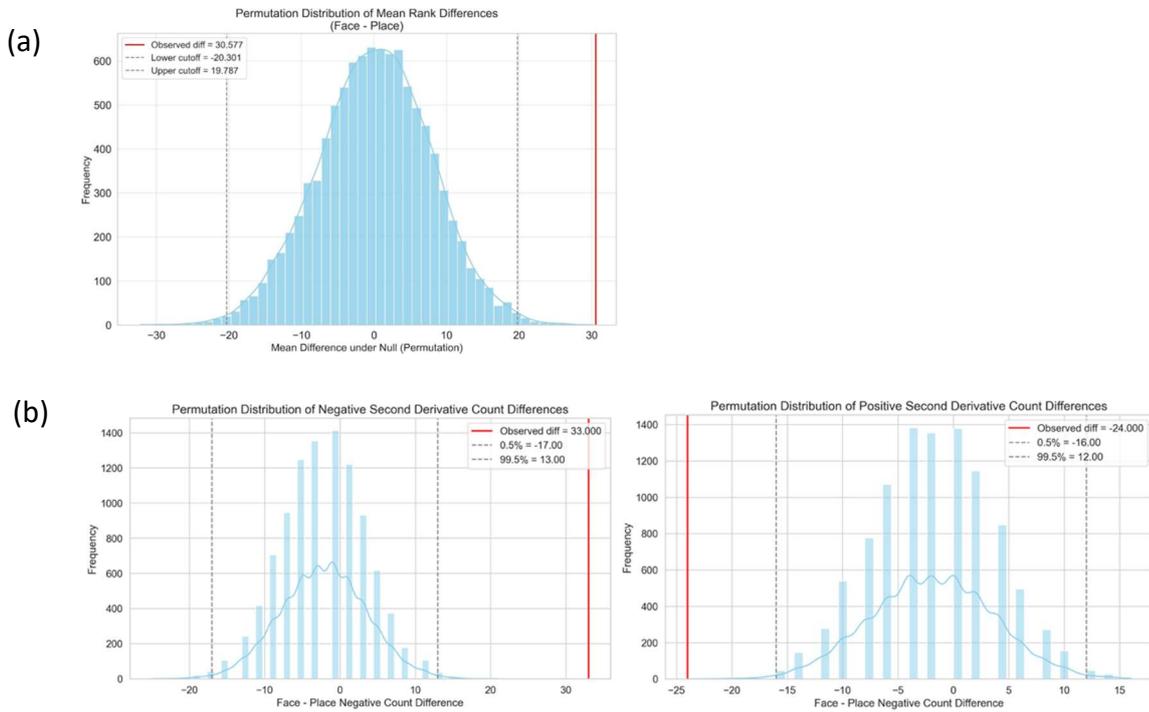

Permutation test results with 10,000 times sample redrawing: (**a**)Permutation distribution of mean S-A rank difference between recruited regions from face and place conditions. The red line indicates the observed difference where face recruits significantly higher S-A rank regions than place on average ($p_{permutation}$ < 0.001). (**b**) Left: the number of concave regions (mean second derivatives < -0.05) from WM face is significantly greater than WM place ($p_{permutation}$ < 0.001). Right: the number of convex regions (mean second derivatives > 0.05) from WM place is significantly greater than WM place ($p_{permutation}$ < 0.001).

Figure S4:

(a)
(b)
(c)
(d)

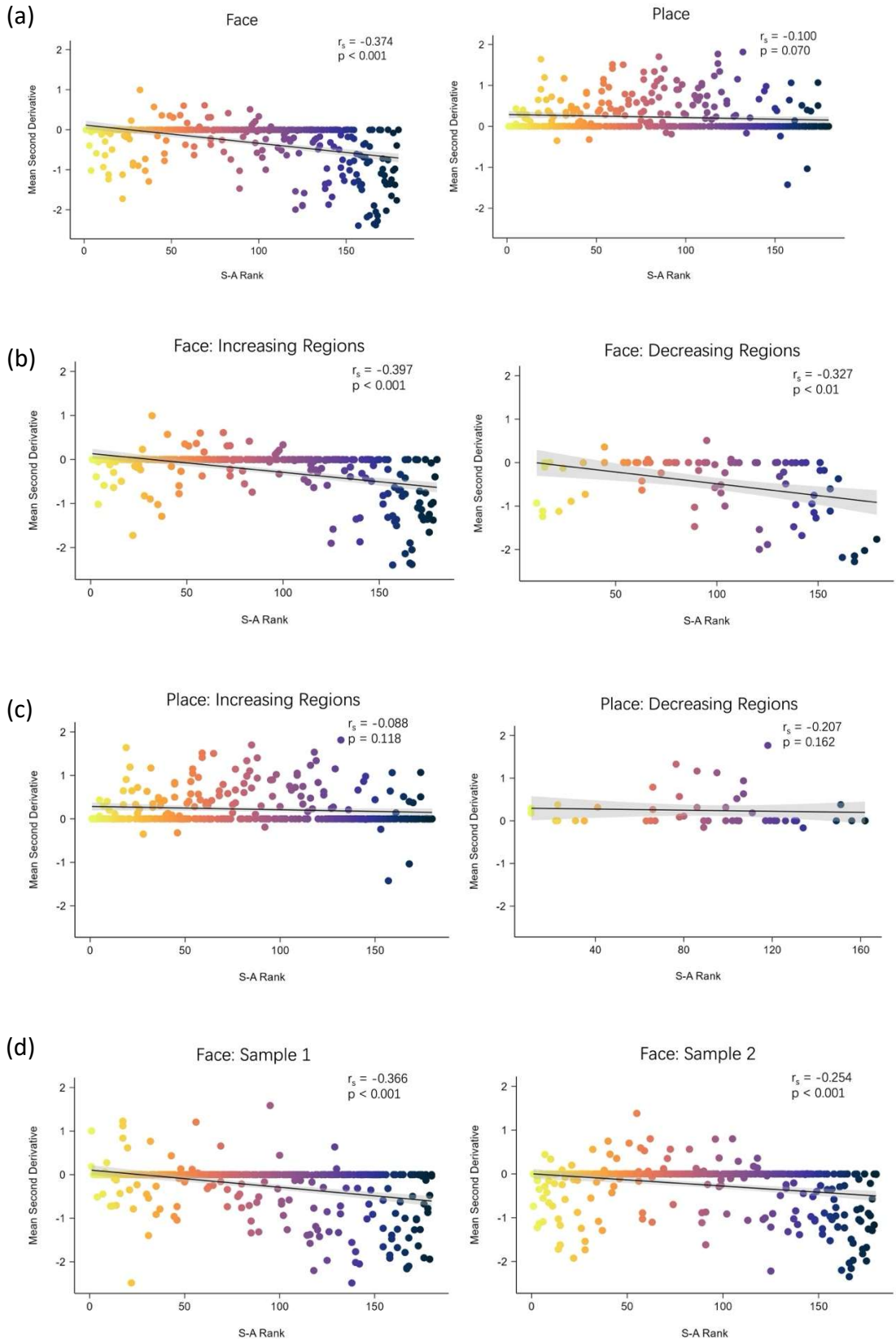

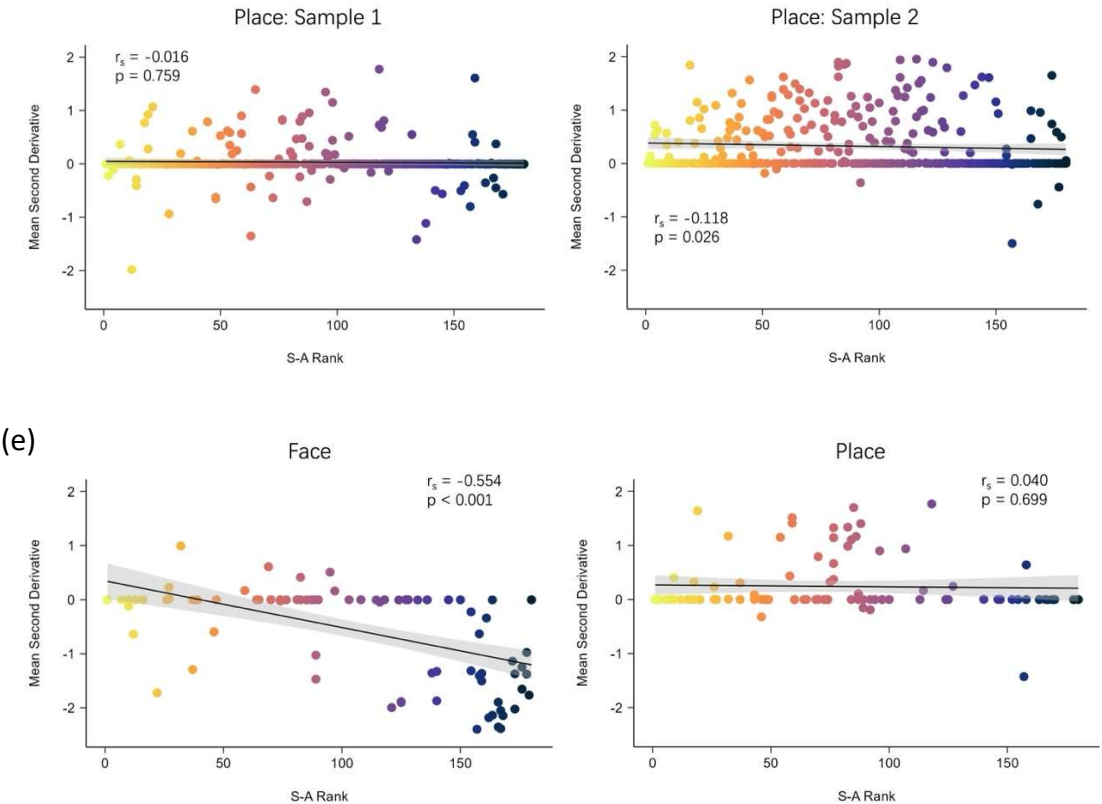

Relationship between cortical hierarchy (S-A ranking) and non-linear activation patterns in each stimulus type, with all 360 cortical parcels from HCP-MMP.

(a)Significant negative correlation between non-linearity and S-A ranking is observed in face task (Spearman's rho = -0.374, p < 0.001), but not in place task (Spearman's rho = -0.100, p = 0.070).

(b)Significant negative correlation between non-linearity and S-A ranking is observed in face task in both regions with positive mean first derivative (Spearman's rho = -0.397, p < 0.001) and negative mean first derivative (Spearman's rho = -0.327, p < 0.01), but not in place task no matter whether it is regions with positive mean first derivative (Spearman's rho = -0.088, p = 0.118) or negative mean first derivative (Spearman's rho = -0.207, p = 0.162).

(c)Two random split sample with the same sample sizes for WM face, both has significant relationship between SA rank and mean second derivative.

(d)Two random split sample with the same sample sizes for WM place, both fail to reach .001 level of significance under spearman regression coefficient.

(e)Significant negative correlation between non-linearity and S-A ranking is observed in regions with significant GAM fit after false discovery rate (FDR) correction in face task (Spearman's rho = -0.554, p < 0.001), but not in place task (Spearman's rho = 0.040, p = 0.699).

Figure S5:

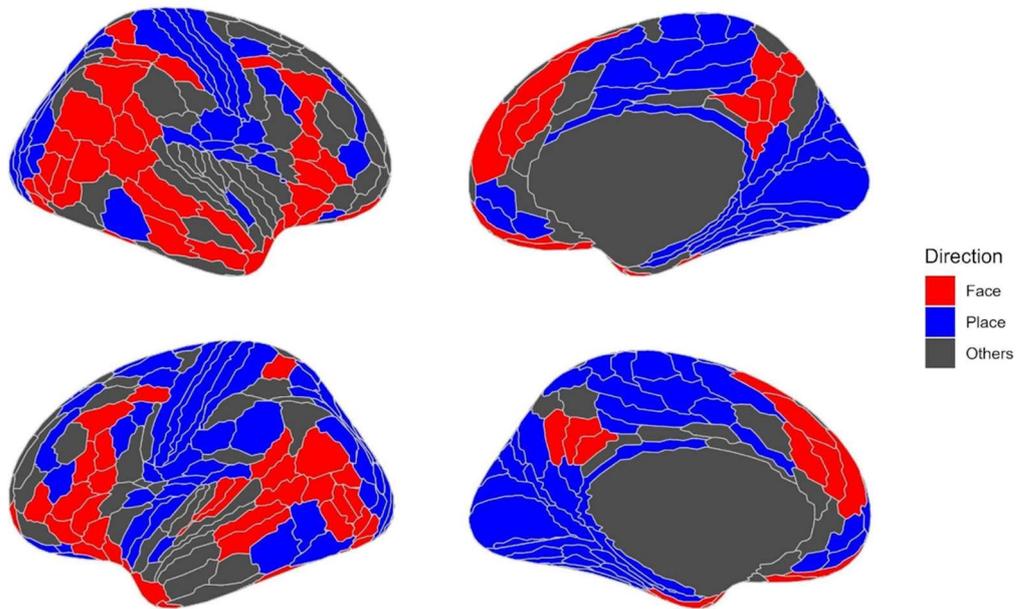

Paired sample t tests of mean activation were performed on each cortical parcel. T-contrast map with FDR correction shows brain regions that have significant higher activation preference for either stimulus. Red regions show activation preference for face stimulus while blue regions show activation preference for place/scene stimulus